\documentstyle[11pt,newpasp,twoside,epsf]{article}
\def\edcomment#1{\iffalse\marginpar{\raggedright\sl#1\/}\else\relax\fi}
\reversemarginpar

\begin{document}
\title{AGN and Starburst Galaxies Seen through Radio Surveys }
 \author{L. Gregorini, I. Prandoni}
 \affil{Universit\'a di Bologna, Italy}
\author{H.R. de Ruiter}
\affil{Osservatorio Astronomico, Bologna, Italy}
\author{P. Parma, G. Vettolani, A. Zanichelli}
\affil{Istituto di Radioastronomia, CNR, Bologna, Italy}
\author{M.H. Wieringa, R.D. Ekers}
\affil{Australia Telescope National Facility, CSIRO, Epping, Australia}

\begin{abstract}
The emergence of a new population of radio galaxies at mJy and sub-mJy levels
is responsible for the change in the slope of the radio source counts.  \\
This population seems to include both star forming galaxies and 
classical (AGN-powered) radio sources, but the relative importance of the two 
classes is still debated. \\
We present results from the ATESP radio survey and its optical
follow-up and show that the fraction of starburst galaxies changes from $\sim$
15$\%$ at fluxes $\ge$ 1 mJy to $\ge$ 50\% at lower fluxes.
\end{abstract}

\section{Introduction}

Recently, deep radio surveys ($S \la 1$ mJy) have shown that
normalized radio counts show a flattening below a few mJy, (see figure 1 
for counts at 1.4 GHz).
This change of slope is generally interpreted as being
due to the presence of a new population of radio sources (the so--called
sub--mJy population) which does not show up at higher flux densities 
(see e.g. Condon 1989), where the counts are dominated by classical radio 
galaxies and quasars.  

The sub-mJy population is essentially composed by low 
luminosity AGN and starburst galaxies, but their relative comtribution is still
not firmely established. 
Unfortunately, due to the long observing times required to reach faint fluxes, 
the existing samples in the sub--mJy region are generally small. 

The identification work and the subsequent spectroscopy are very demanding 
in terms of telescope time.
Typically, no more than $\sim 50-60\%$ of  the radio sources in sub--mJy 
samples have been identified on optical images, even though in the $\mu$Jy 
survey in the Hubble Deep Field  and in SSA13 80$\%$ of the 111
radio sources have been identified (Richards et al. 1999). 
On the other hand, the typical fraction of spectra available is only 
$\sim 20\%$. The best studied sample is the Marano 
Field, where 50$\%$ of the sources have spectra (Gruppioni et al. 1999a),
which permitted the determination of spectral type and redshift for 29 objects. 
To alleviate the identification work, regions with 
deep photometry already available (possibly multicolor) provide a significant 
advantage.  
The region we have selected fulfills these requirements at least partially.

\begin{figure}
\epsfxsize=10cm
\epsfysize=10cm
\epsfbox{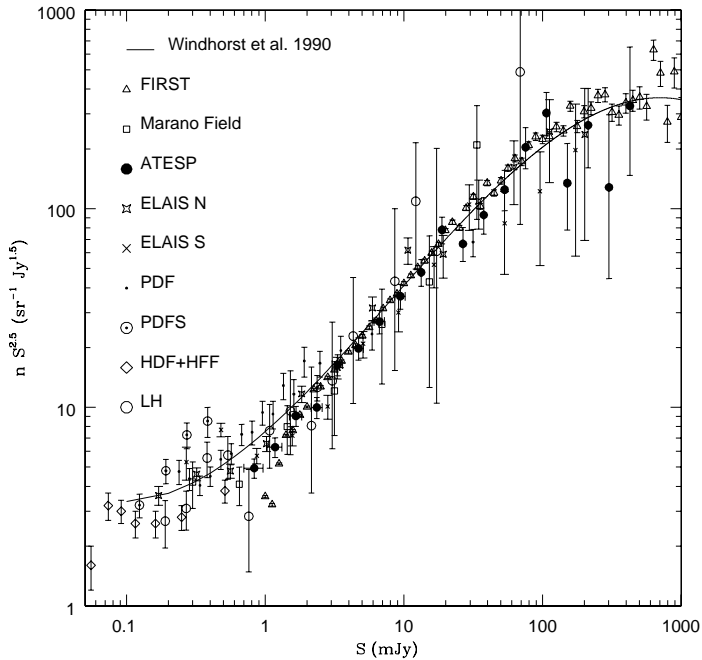}
\caption{Normalized 1.4 GHz differential source counts for different samples.
The ATESP source counts are indicated by filled circles.
The fit obtained by Windhorst et al. (1990) is also indicated (solid line). }
\end{figure}

\section{The ATESP survey}

Vettolani et al. (1997) made a deep redshift survey near the SGP by studying 
photometrically and spectroscopically nearly all galaxies down to $b_J 
\sim$ 19.4. The survey, yielding 3342 redshifts (Vettolani et al. 1998), 
has a typical depth of $z=0.1$ with 10$\%$ of the objects at
$z>0.2$ and is $90\%$ complete.

In the same region a radio survey has been carried out with the ATCA 
(Australia Telescope Compact Array) at 1400 MHz (Prandoni et al. 2000a,b). 
It consists of 16 radio mosaics with $\sim 8\arcsec \times 14\arcsec$
resolution and uniform sensitivity ($1 \sigma$ noise 
level $\sim$79~$\mu$Jy) over the whole area of the ESP redshift survey 
($\sim 26$ sq.~degrees at $\delta \sim -40\deg$).  
We detected $2960$ distinct radio sources down to a flux density limit of 
$\sim 0.5$ mJy ($6\sigma$), 1403 being sub--mJy sources.

We used the ATESP catalogue to derive the differential ATESP source 
counts down to 0.70 mJy (Prandoni et al. 2000c).
These counts are compared with the most updated previous determinations
 at 1.4 GHz (see Fig.~1). Also shown is the 
interpolation determined by Windhorst et al. (1990) from a collection of 
$10,575$ radio sources belonging to 24 different surveys at 1.4 GHz, 
representing the state of the art at that time (solid line). 
 There is consistency between the ATESP counts and those obtained by 
other recent surveys, with the exception of the Phoenix Deep 
Survey (PDF and PDFS, Hopkins et al. 1998), whose counts at $S\geq 0.7$ mJy 
are systematically higher than the ATESP counts (and also higher than 
the counts derived from the other surveys presented in the figure). 

The ATESP counts are in very good agreement with the FIRST counts 
(White et al. 1997), which are the most accurate available today over the 
flux range 2--30 mJy. 
The ATESP survey, on the other hand, provides the  best determination of the 
counts 
at fainter fluxes ($0.7<S<2$ mJy), where the FIRST becomes incomplete. 
The ATESP counts can thus provide an useful observational constraint
on the evolutionary models for the mJy and sub--mJy populations. 

\begin{figure}
\epsfxsize=10cm
\epsfysize=10cm
\epsfbox{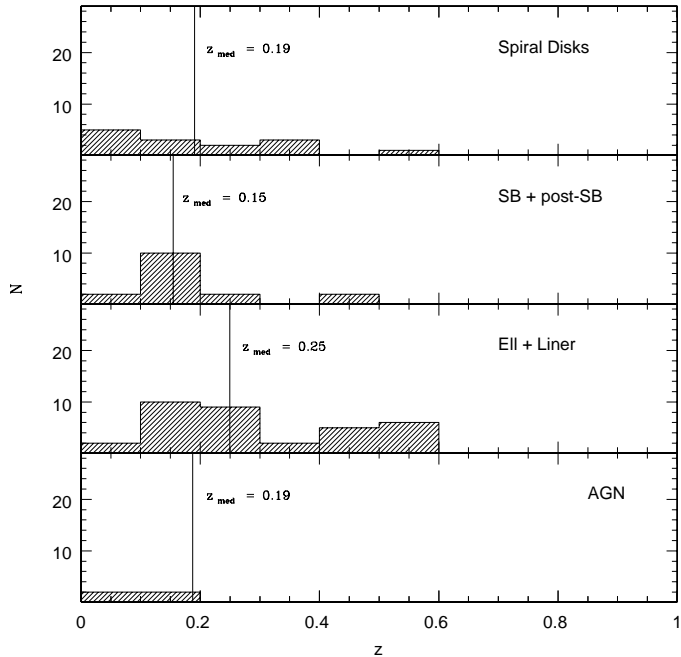}
\caption{Redshift distribution for the different spectral classes.}
\end{figure}

\section{The ATESP-EIS Sample}

In the same region lies the EIS (ESO Imaging Survey, Nonino et al. 1999)
Patch A (3.2 sq. degr.), consisting of deep images in the I band out of which
a galaxy catalogue to I=22.5 has been extracted. 
For 218 out of 384 radio sources ($S \geq 0.5$ mJy) in 2.97 square 
degrees of the EIS-A  an optical identification was found; we obtained spectra
at the ESO 3.6m telescope of a complete sample of 69 galaxies brighter 
than I = 19.0. The high signal 
to noise ratio of these spectra permits an unambiguous spectral classification 
of the whole sample. We divided the galaxy-identifications into several
standard groups; this classification, based on spectral type, essentially
distinguishes between AGNs, which, for weak radio sources, are mostly
elliptical galaxies without strong emission lines (although also a few
Seyfert type spectra are present), and starburst galaxies (McCall et al.
1985), which are often spiral galaxies in a particularly active phase of
star formation. As expected, we found a number of these starburst
galaxies, which are characterized by high excitation $HII$ region-like
spectra, but also some post-starbursts.  The latter have either strong 
$H\delta$
absorption on top of a K-type spectrum without emission lines, or show
$H\alpha$ and $H\beta$ in emission and  higher order lines (from
$H\delta$ on) strong but in absorption. In addition some spectra are
characteristic of galaxy disks of late type spirals: in those cases some
emission is present ([OIII],[OII],$H\alpha$ and NII) but at a lower level
than in the starburst galaxies. Finally, one typical LINER (Heckman 1980)
was found.

Figure 2 shows the redshift distribution of the different spectral
classes. The mean redshift distribution of the whole sample is z=0.20; 
in particular
starburst and post-starburst galaxies are nearer than ellipticals, in good
agreement with the results from FIRST (Magliocchetti et al. 2000).

\section{The mJy and sub-mJy Population}

\begin{table}[t]
\small
\caption[]{Composition of faint radio population} 
\begin{tabular}{lrrrrrr}
\multicolumn{7}{l}{}\\
\hline
\multicolumn{1}{l}{Type}  & \multicolumn{2}{c}{All} &
\multicolumn{2}{c}{$S<1$ mJy} & 
\multicolumn{2}{c}{$S\geq 1$ mJy} \\
\multicolumn{1}{c}{}  
& \multicolumn{2}{c}{\footnotesize{$N \; \; \; \; \; \;  \; \;  \%$}} &
\multicolumn{2}{c}{\footnotesize{$N \; \; \; \; \; \;  \; \;  \%$}} & 
\multicolumn{2}{c}{\footnotesize{$N \; \; \; \; \; \;  \; \;  \%$}} \\
\hline 
  & & & & & & \\
 Ell+liner    & 34 & (50\%) & 7 & (29\%) & 27 & (60\%)    \\
 AGN    & 5 & (7\%) & 1 & (4\%) & 4 & (9\%)    \\
 Spiral disks & 14 & (20\%) & 6 & (25\%) & 8 & (18\%) \\
 SB + post-SB & 16 & (23\%) & 10 & (42\%) & 6 & (13\%) \\
 All & 69 & & 24 & (35\%) & 45 & (65\%) \\
  & & & & & & \\
\hline\hline 
\end{tabular}
\end{table}

The faint radio source composition resulting from the ATESP-EIS spectroscopic 
sample classification is presented in Table~1, where
sub-mJy and mJy regimes have been considered separately. We notice that
the good quality of the spectra allowed us to classify all objects
(in previous spectroscopic studies of sub-mJy
samples about 15-20$\%$ of the objects were not classified due to poor
spectroscopy). \\
Our data clearly show that the AGN contribution  does not 
significantly change going 
to fainter fluxes (from 9\% to 4\%), that early-type galaxies largely 
dominate (60\%) the mJy 
population, while star-formation processes become important in the sub-mJy 
regime: SB and post-SB galaxies go from 13\% at $S\geq 1$ mJy to 42\% at 
$S<1$ mJy.  
Nevertheless, at sub-mJy fluxes, early-type galaxies still constitute a 
significant fraction (29\%) of the whole population. As shown in Fig.~3
where we plot the radio flux 
densities against the I magnitudes for the whole sample,
this seems to be 
particularly true going to fainter magnitudes. In fact at $I>18$ 40\% ($2/5$) 
of the sub-mJy sources are early-type galaxies (circles) 
and the fraction of starburst galaxies (stars) in only 20\% ($1/5$).  

\begin{figure}
\epsfxsize=10cm
\epsfysize=10cm
\epsfbox{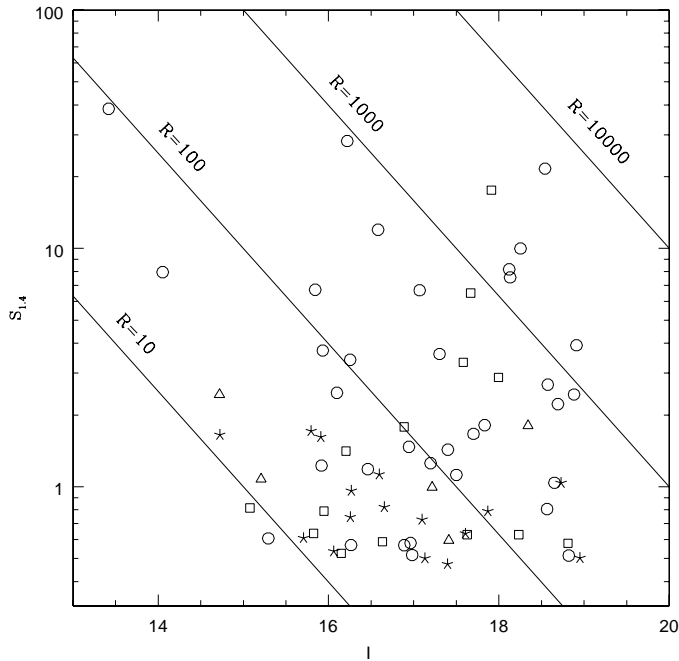}
\caption{Flux density versus I magnitude; lines represent constant radio to 
optical ratio, defined as: $R~= S_{1.4GHz} \times 10^{0.4(m-12.5)}$. Symbols 
represent the different classes of objects: Ell. + Liners (circles), 
spiral disks (squares), SB + post-SB (stars) and AGN (triangles).}
\end{figure}

The latter result, even though based on a very small number of objects,
is in agreement with the result obtained from the analysis 
of the Marano Field sub-mJy sample (Gruppioni et al. 1999a), and suggests that 
star-forming galaxies dominate the sub-mJy population only at bright 
magnitudes. Deeper spectroscopy for the ATESP-EIS sample will be crucial 
in order to verify this indication on a reliable statistical basis. \\
Fig.~3 also indicates a possible physical interpretation of this 
result: star-forming galaxies are characterized by smaller radio 
to optical ratios ($10<R<100$), that is have weaker intrinsic radio emission,
than early-type galaxies ($R\geq 100$). If this behaviour holds going to
fainter fluxes, a larger fraction of star-forming galaxies is expected
in $\mu$Jy samples. This hypothesis is supported by the study
of the $\mu$Jy sources in the Hubble Deep Field, the majority of which seem
to be associated with star-forming galaxies (Richards et al. 1999). 
 
\section{Perspectives}

In order to improve our knowledge of the sub-mJy population it is crucial
to obtain deeper optical images and/or fainter radio samples. 
The first step will be to analyze the UBVRI images of DEEP-1 field
(EIS survey) which overlaps a subregion of the ATESP survey.
At a limiting magnitude of about I=26 we estimate to indentify 70-80$\%$ of
the 135 radio sources present in the field. Furthermore the same region
will be observed at 5 GHz with the ATCA; radio spectral index and radio 
source structure  will provide important clues on the nature of sub-mJy 
sources and put strong observational constraints on evolutionary models.    

The next step is to have a $\mu Jy$ sample with high quality spectroscopy;
the VLA data at 1.4 GHz for the deep field of the VIRMOS spectroscopic
survey (Le Fevre, these proceedings) fulfill these requirments.

\end{document}